\documentclass[conference]{IEEEtran}
\IEEEoverridecommandlockouts
\usepackage{cite}
\usepackage{amsmath,amssymb,amsfonts}
\usepackage{algorithmic}
\usepackage{float}
\usepackage{graphicx}
\usepackage{xcolor}
\usepackage{tabularx}
\usepackage{multirow}
\usepackage{lipsum}
\usepackage{enumitem}
\usepackage{booktabs}
\usepackage{ragged2e}
\graphicspath{ {figures/} }
\usepackage{filecontents}
\newcolumntype{Y}{>{\RaggedRight\arraybackslash}X} 
\def\BibTeX{{\rm B\kern-.05em{\sc i\kern-.025em b}\kern-.08em
    T\kern-.1667em\lower.7ex\hbox{E}\kern-.125emX}}
\begin{document}

\title{Comparative Study of ZF, LMS and RLS Adaptive Equalization for alpha-mu Fading Channels}

\author{
  Rayyan Abdalla\\
  Department of Electrical, Computer and Energy Engineering\\
  Arizona State University \\
  \texttt{rsabdall@asu.edu} \\
}

\maketitle

\begin{abstract}

Wireless communication systems generally endure severe fading and interference caused by the time-dispersive channel.  Major transmission distortion is produced by channel multipath propagation and overlap of subsequent symbols. To counteract channel response, equalization techniques are employed to operate on channel output and recover transmitted signal at the receiver. This work investigates equalization of channels undergoing $\alpha-\mu$ fading distribution, which models multipath propagation in millimeter Wave (mmWave) and sub-Terahertz (sub-THz) frequencies bands. Three equalization algorithms, which are non-adaptive Zero-Forcing (ZF), adaptive Least-Mean-Square (LMS) and adaptive Recursive-Least-Square (RLS) are implemented and addressed in terms of Bit Error Rate (BER), Convergence speed and implementation complexity. Monte Carlo simulations are then carried out to compare algorithms and assess performance by varying multiple parameters such as training lengths, channel order, equalization taps and diverse fading conditions. 

\end{abstract}


\section{Introduction}\label{introduction}
Modern wireless communication technologies are required to achieve data transmission at high rates within constrained power resources and complexity. These requirements impose interference mitigation and reliable operation in noise environments. As wireless broadband applications increasingly demand large bandwidth, several challenges are encountered at higher frequency bands in terms of multipath (small-scale) fading and severely attenuated communication channel. The resulting dispersion of channel frequency response causes smeared transmission and inseparable transmitted symbols, which is referred to as inter-symbol interference (ISI). One possible technique to reduce the distortion produced by the multipath propagation is channel equalization at the receiver.  A channel equalizer is a digital filter that inverts the response of the multipath channel in such a way that the BER is kept low at reasonable Signal to Noise Ratio (SNR). 
The function of a channel equalizer quantifies the communication channel at the receiver to recover the transmitted signal. While a simple solution, ZF equalization for instance \cite{krovvidiperformance,sahoo2015channel}, suggests implementing an inverse filter with a static (non-adaptive) transfer function, such equalizer performance is not effective since it neglects channel variations and noise effect. Therefore, adaptive equalizers are suited to track the time-varying channel response and compensate for additive noise in transmission medium.  Two common linear adaptive equalizers are based on LMS and RLS algorithms. The LMS algorithm is simpler and more robust than the RLS algorithm, however, it has slower convergence compared to that of the RLS algorithm \cite{mosleh2010combination}. The choice between LMS and RLS approaches is dependent on many factors such as channel coherence time, hardware complexity and transmitted signal characteristics, which in turn determines algorithm stability. 
Since equalizer design depends on the multipath channel statistics, it is essential to comprehensively model the fading experienced by the communication channel to address the degree of attenuation and ISI. Main channel dispersion sources include path propagation loss, shadowing effect and small-scale fading. While shadowing and path loss can be modeled deterministically, small-scale fading is stochastic and dominant. Several fading distributions were proposed to model multipath channel envelop, such as Rayleigh, Rice, and Nakagami-m distributions. Although widely adopted, these statistical models capture channel behavior at certain conditions, particularly at lower frequencies up to the sub-6 GHz band. As new mobile communication systems are envisioned to utilize available spectrum at millimeter Wave (mmWave) and sub-THz bands, corresponding fading at these bands is modeled by means of a generalized Gamma-based distribution known as $\alpha-\mu$ distribution \cite{boulogeorgos2019analytical,papasotiriou2021new,fraidenraich2006alpha}. That is, this distribution provides theoretical characterization of coefficients magnitude of the channel impulse response to be equalized. 
The problem of channel equalization is a frequently addressed topic in wireless communication research. It is observed that most efforts are in the direction of enabling adaptive equalizers. The work in \cite{sahoo2015channel} implemented an LMS-modified ZF algorithm for Rayleigh fading channels. A similar analysis was introduced using ZF technique aided with interference cancellation by the authors in \cite{krovvidiperformance}. A joint channel estimation approach combining LMS and RLS algorithms was proposed by \cite{mosleh2010combination}. The authors in \cite{sharma2014performance} analyzed general ISI channel using ZF, LMS and RLS algorithms and provided interesting results in terms of recommended training symbols and BER performance. The work in \cite{khajababu2015channel} investigated modulation effect on equalizing Rayleigh distributed channel using LMS and RLS algorithms. Moreover, some recent efforts related to high-frequency channel equalization with time-reversal signaling and hardware impairment mitigation were conducted by \cite{gharouni2017performance} and \cite{sha2021channel} respectively. While most tutorials focus on addressing single or two equalization algorithms for a specific fading distribution, this work presents a comprehensive analysis and performance assessment of ZF, LMS and RLS pilot-aided equalization algorithms designed for a generalized multipath communication channel following  $\alpha-\mu$ distribution in the presence of additive white Gaussian noise (AWGN). Note that blind equalization is not covered in this article. The paper organization is follows: section \ref{sec:channel_equalization} illustrates the theory of channel equalization and implementation of the three algorithms. Section \ref{sec:sys_model} presents the proposed baseband system model structured for simulations. Simulation results are discussed in section \ref{sec:sim} and concluding remarks are drawn in section \ref{sec:conclusion}.  

\section{Channel Equalization} \label{sec:channel_equalization}
\begin{figure}[htb] 
\centering
\includegraphics[width = 9 cm,height = 4.7 cm]{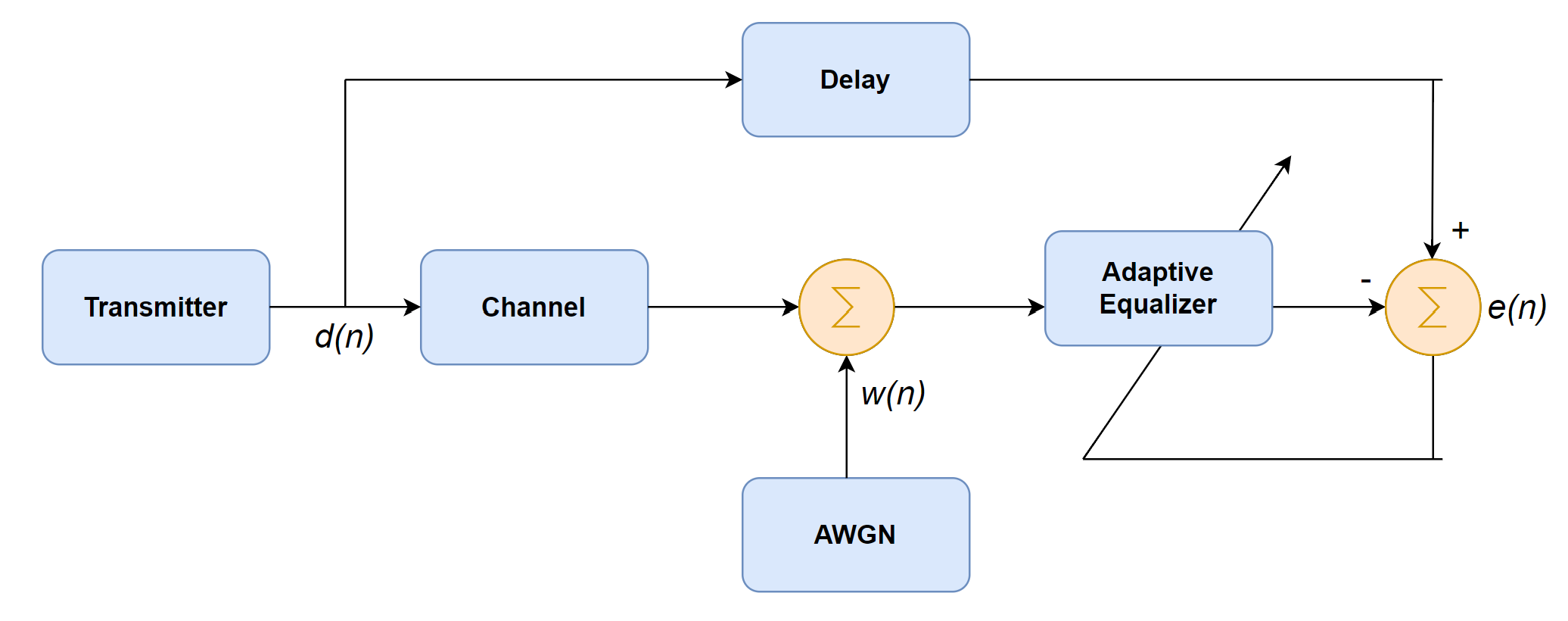}
\caption{Block diagram of adaptive channel equalization}
\label{fig1}
\end{figure}

In a communication system, the multipath channel acts a time-dispersive band-limited filter that distorts the transmitted signal, which in turns causes bit errors at the receiver. A channel with ISI can be modeled as a multi-tap filter with each tap corresponding to a multipath component. The idea of channel equalization is to introduce a digital filter at the receiver that takes distorted transmitted signal as an input and produces a delayed version of the originally transmitted signal. Channel equalization can be implemented statically, meaning equalizer impulse response is made constant as the inverse of estimated communication channel impulse response, or adaptively, where equalizer impulse response continuously adapts with the varying channel impulse response. Static equalizers, such as ZF equalizers, are normally pre-set linear filters that use peak distortion criterion to estimate equalizer coefficients \cite{krovvidiperformance,sahoo2015channel,sharma2014performance}. Adaptive equalization, on the other hand, is based on transmitting a signal within channel coherence time and known to both transmitter and receiver, which in turn gets distorted by the multipath channel and AWGN. The corresponding distorted signal is then utilized by the receiver to adjust equalizer tabs by attempting to produce a delayed version of the originally transmitted signal. Once equalizer taps are set, the predicted filter coefficients can then be used to mitigate both channel dispersion and noise affecting subsequent transmissions in the same coherence period. Both LMS and RLS algorithms are popular adaptive system identification algorithms that are used for equalizing communication channels \cite{mosleh2010combination,khajababu2015channel}. Figure \ref{fig1} shows a block diagram of adaptive channel equalization.  

\subsection{Zero-Forcing Equalization}
The method of ZF utilizes estimated ISI channel to design suitable linear filter with transfer function inverse to channel impulse response. The ZF equalizer forces ISI components to zero while ignoring the additive noise in the channel. While ZF equalizer operates ideally in a noise-free case, it is less efficient in practical conditions since it may significantly amplify the noise if the channel response has spectral nulls. Moreover, finite-length channel impulse responses require infinitely long ZF impulse response \cite{sharma2014performance}. Therefore, ZF equalization is restricted to certain applications where ISI components are dominant, and the additive noise is negligible.
Let $H_{chn}(f)$ be the channel transfer function with phase $\theta_{chn}(f)$, and $H_{equ}(f)$ be the equalizer transfer function. Both functions are related according to
\begin{equation}
    H_{chn}(f) = |H_{chn}(f)|e^{j\theta_{chn}(f)}, \hspace{0.1in} |f|\leq W
\end{equation}
\begin{equation}
    H_{equ}(f) = \frac{1}{H_{chn}(f)}=\frac{1}{|H_{chn}(f)|}e^{-j\theta_{chn}(f)},\hspace{0.05in} |f|\leq W
\end{equation}
\par

\subsection{Least-Mean-Square Equalization}
The adaptive LMS algorithm is a system identification gradient decent method that minimizes the Mean Square Error (MSE) between desired filter response and the adaptive filter output. The algorithm iteratively estimates error gradient vector and uses it to adjust filter weights, attempting to converge on optimal Weiner solution \cite{sharma2014performance,khajababu2015channel}. The LMS algorithm is widely adopted in several applications due to low implementation complexity and robustness. For a desired filter response $d(n)$, adaptive filter output $y(n)$, input vector \textbf{$x(n)$}  and error signal $e(n)$, and weights vector \textbf{$b(n)$}, the LMS algorithm is executed in two steps:
\begin{itemize}
    \item{Estimation process: this involves computing the error within iteration \textit{n} as a function of desired signal and generated filter output:} 
    \begin{equation}
        e(n) = d(n)-y(n)
    \end{equation}
    This error is used to calculate the gradient vector:
    \begin{equation}
        \textbf{$\nabla$}(n)=-e(n)\textit{$\textbf{x}$}^*(n)
    \end{equation}

    \item Adaptation process: this involves automatic adjustment of the adaptive filter weights based on gradient vector, i.e., the estimated error:
     \begin{equation}
       \textit{ \textbf{b}}(n+1)=\textit{\textbf{b}}(n)-\eta \textit{\textbf{$\nabla$}}(n)
    \end{equation}
     \begin{equation}
       \textit{\textbf{b}}(n+1)=\textit{\textbf{b}}(n)+\eta e(n)\textit{\textbf{x}}^*(n)
    \end{equation}
    Where $$\eta$$ is the step size that controls convergence speed and thus directly determines algorithm stability. Note that all vectors are of size $$(L+1)×1$$ where $$L$$ is the adaptive filter order. 
    In a channel equalization case and when adapting equalizer taps, the pilot signal represents the desired filter response while the distorted pilot by the channel represents filter input. The equalizer filter is then desired to cancel channel effect and produce the originally transmitted signal at the receiver. 

\end{itemize}

\subsection{Recursive Least Square Equalization}
The RLS method is a recursive algorithm that is similar to LMS in terms of minimizing MSE and implementation processes. However, a weighted RLS algorithm minimizes a least square cost function corresponding to input signal. An RLS filter whitens the input data by evaluating and utilizing inverse correlation matrix within every iteration to update filter weights. The rate of RLS convergence is several times faster than that of LMS, which makes the algorithm a great fit for fast fading communication channel. However, RLS involves high computation complexity and may encounter stability challenges during the calculation of inverse autocorrelation function \cite{mosleh2010combination,khajababu2015channel,sharma2014performance}.  

For a desired filter response $d(n)$, input vector \textit{$\textbf{x}(n)$}  and error signal $e(n)$, and weights vector \textit{$\textbf{b}(n)$}, time-varying gain vector \textit{$\textbf{k}$(n)}, forgetting factor $\gamma$, and input autocorrelation matrix \textit{$\hat{\textbf{R}}$}(n), the RLS algorithm, like MSE, is executed in two steps:
\begin{itemize}
    \item Estimation process: which involves estimating error and gain vector as follows:
    \begin{equation}
        \textit{$\textbf{k}$(n)} = \frac{\gamma^{-1}\hat{\textbf{R}}^{-1}(n-1)\textbf{x}(n)}{1+\gamma^{-1} \textbf{x}^H(n) \hat{\textbf{R}}^{-1}(n-1)\textbf{x}(n)}
    \end{equation}

    \begin{equation}
        e(n) = d(n) -  \textbf{b}^H(n-1)  \textbf{x}(n)
    \end{equation}
    With initial conditions $\hat{\textbf{R}}^{-1}(0) = I$ , $\textbf{b}(0)=0$, $0<\gamma<1$

    \item	Adaptation process: which involves filter taps updates as follows:
    \begin{equation}
       \textit{\textbf{b}}(n)=\textit{\textbf{b}}(n-1)+ \textit{\textbf{k}}(n)e^*(n)
    \end{equation}

    \begin{equation}
       \hat{\textbf{R}}^{-1}(n)=\gamma^{-1}\hat{\textbf{R}}^{-1}(n-1)-\gamma^{-1} \textit{$\textbf{k}$(n)}  \textbf{x}^H(n) \hat{\textbf{R}}^{-1}(n)
    \end{equation}
    
\end{itemize}
In adaptive channel equalization case and when adapting equalizer taps, the pilot signal represents the desired filter response while the distorted pilot by the channel represents filter input. The equalizer filter is then desired to cancel channel effect and produce the originally transmitted signal at the receiver. 
\section{General System Model}\label{sec:sys_model}
The communication system model employed in this study include Binary Phase Shift Keying (BPSK) modulated complex baseband system with prominent sections of transmitter, multipath channel, and receiver. Assuming normalized energy, the transmitter transmits a signal $u_t$, which propagates through ISI channel $h$ with $K$ multipath components, i.e. channel taps. Note that with $K=1$, the multipath channel is assumed flat with a constant gain effect. The receiver signal $u_r$  consists of multipath components combined with AWGN represented by $w$. The system equation at a given time $n$ is given by:
\begin{equation}
    u_r(n)= \sum_{k=1}^{K-1} u_t(n) h(n-k)  + w(n)\
\end{equation}

The channel taps follow $\alpha-\mu$ fading distribution, which comprises a Gamma-based general distribution. This fading distribution has been proven to successfully describe small-scale fading in Radio Frequency (RF), mmWave and sub-THz channels \cite{boulogeorgos2019analytical,papasotiriou2021new,gharouni2017performance}. Moreover, with a certain parameter setting, $\alpha-\mu$ Probability Distribution Function (PDF) yields many well-known fading distributions, such as Rayleigh, Weibull, Nakagami-m and one-sided Gaussian, which makes it a great fit for fading modeling at various channel conditions. The PDF and Cumulative Density Function (CDF) of $\alpha-\mu$ distribution are shown below:

\begin{equation}
    f(x) = \frac{\alpha \mu^{\alpha} \left( \frac{x}{\beta}\right)^{\alpha \mu -1 } \exp{\left( -\mu \left( \frac{x}{\beta}\right)^{\alpha} \right)}}{\beta \Gamma(\mu)}
\end{equation}

\begin{equation}
    F(x) = 1 - \frac{\Gamma \left( \mu, \mu  -\mu \left( \frac{x}{\beta}\right)^{\alpha} \right)}{\Gamma(\mu)}
\end{equation}
\begin{equation}
    \beta = \sqrt[\alpha]{E\left( X^\alpha \right)}
\end{equation}

\begin{figure}[htb] 
\centering
\includegraphics[width = 9 cm,height = 6 cm]{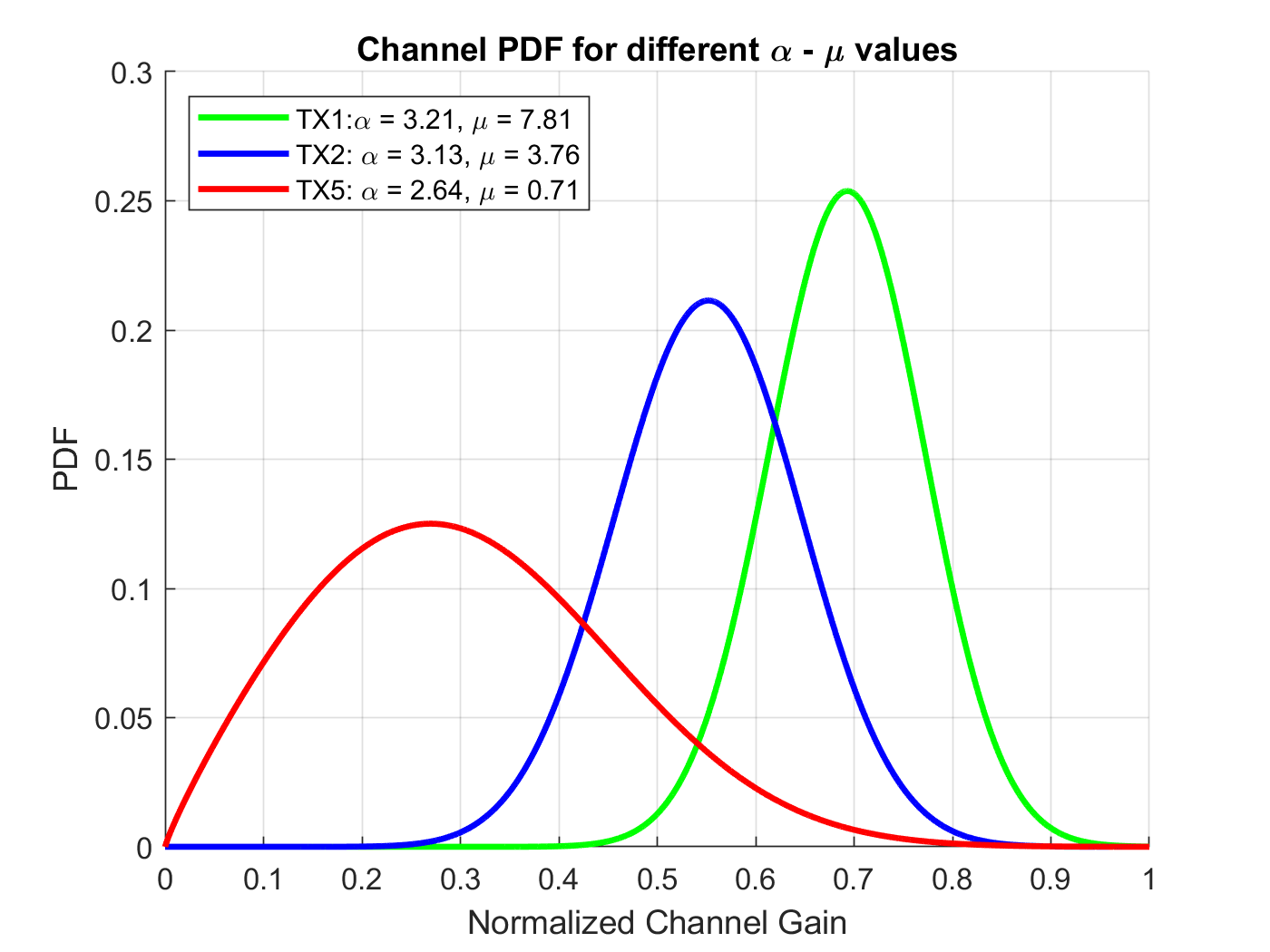}
\caption{Channel PDF for different $\alpha-\mu$ values}
\label{fig2}
\end{figure}

Where $\Gamma (.)$ is the complete Gamma function. In order to determine practical $\alpha-\mu$ parameter values that resemble realistic communication channel, one can follow the estimations of \cite{papasotiriou2021new}. These findings are based on empirical channel measurements for both Line-of-Sight (LOS) and non-LOS sub-THz transmission links. From figure (1) and Table (1) of \cite{papasotiriou2021new}, three RX-TX pairs are considered for study, which are RX-TX1 with $\alpha$$=3.21$,$\mu$$=7.81$ (LOS), RX-TX2 with $\alpha$$=3.13$,$\mu$$=3.76$ and RX-TX5 with $\alpha$$=2.64$,$\mu$$=0.71$ (non-LOS). The PDFs of the corresponding three RX-TX pairs are shown in figure \ref{fig2}.

Now that system parameters are specified, simulations are carried out using MATLAB to investigate the performance of ZF, LMS and RLS equalization algorithms. The first stage of simulation is based on transmitting the pilot signal to adjust equalizers coefficients. Convergence speed is analyzed in this stage. The second simulation stage is concerned with running Monte Carlo simulations by generating several streams to pass through the multipath channel. These streams are processed at the receiver to mitigate channel and noise distortion and therefore, performance is assessed in term of BER curves. 

\section{Experiments and Results}\label{sec:sim}
This section presents computer simulation results for ZF, LMS and RLS equalization algorithms. Each algorithm is analyzed in terms of BER performance, adaptive algorithms are moreover addressed in terms of convergence speed, behavior according to varying channel parameters and optimal lengths of training and equalization taps. That is, a major analysis objective is to illustrate how $\alpha-\mu$ distributed multi-tap channel is mitigated with channel equalization.

\subsection{Zero-Forcing Performance}
\begin{figure}[htb] 
\centering
\includegraphics[width = 9 cm,height = 6.6 cm]{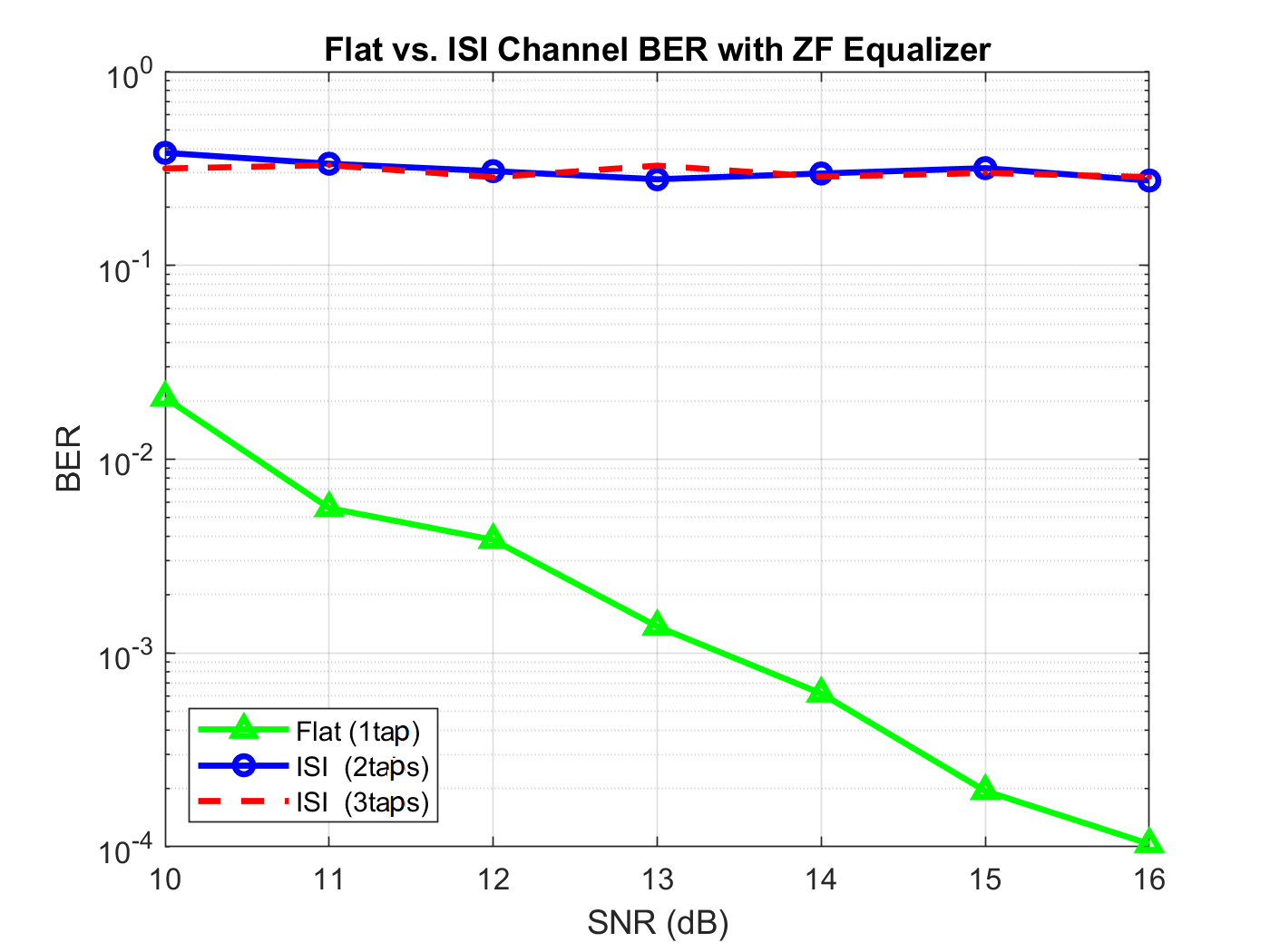}
\caption{ZF equalizer performance on flat vs. ISI channel}
\label{fig3}
\end{figure}

The algorithm is examined in terms of ability to equalize flat fading (single channel taps) and ISI channel (2-3 channel taps). The LOS RX-TX1 pair of figure (2) and Table (1) in \cite{papasotiriou2021new} is assumed for setting channel parameters. Generated streams for Monte Carlo simulations are of length 1000 symbols. The equalizer impulse response is the inverse of channel transfer function. The BER performance for flat versus ISI channel is shown in figure \ref{fig3}. 

For a reasonable range of SNR, the ZF equalizer error probability is much lower for flat fading channel compared to multi-tap channel. Recall that ZF algorithm ignores the additive noise, meaning that it is only effective when ISI distortion is dominant under high SNR values. 

\subsection{Least-Mean-Square Performance}
The algorithm is evaluated in terms of equalizing flat and ISI channel and equalizing channel under several $\alpha-\mu$ parameter settings. Different training lengths and equalization orders and are tested to optimize BER performance. A minimum number of 10000 streams of length 1000 symbols are generated for Monte Carlo simulations. The algorithm is run 100 times to estimate MSE and average BER. For suitable convergence and minimal misalignment, the learning rate, $\eta$ is selected 0.04. Unless otherwise specified, all BER and convergence curves are obtained assuming RX-TX1 pair of figure \ref{fig2} with a training length of 1000 symbols and 16-tap equalizer. The analysis is conducted according to the following criterions:

\subsubsection{Flat vs. ISI Channel Equalization}
\begin{figure}[htb] 
\centering
\includegraphics[width = 9 cm,height = 6.6 cm]{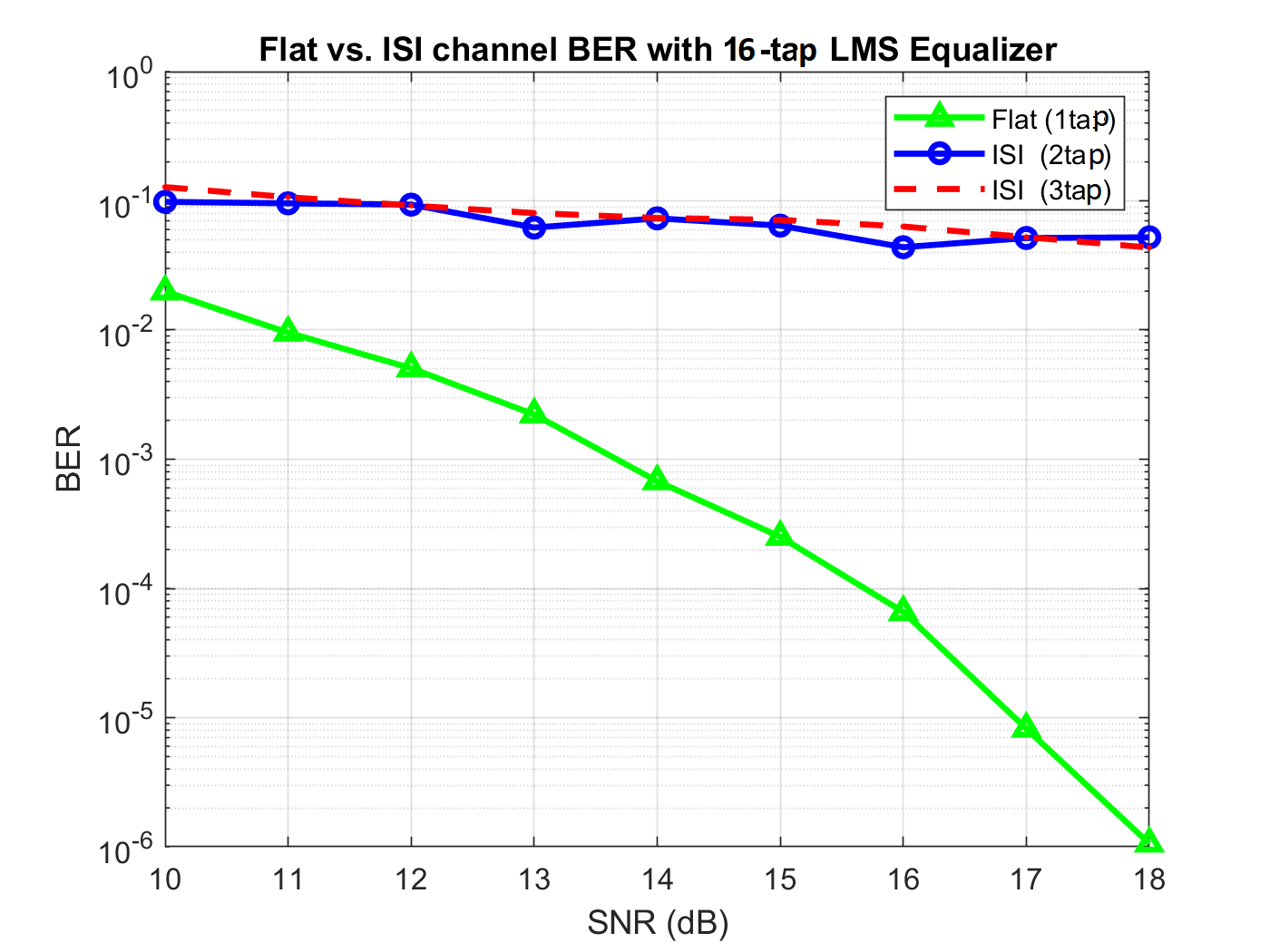}
\caption{LMS equalizer performance on flat vs. ISI channel}
\label{fig4}
\end{figure}

\begin{figure}[htb] 
\centering
\includegraphics[width = 9 cm,height = 6.6 cm]{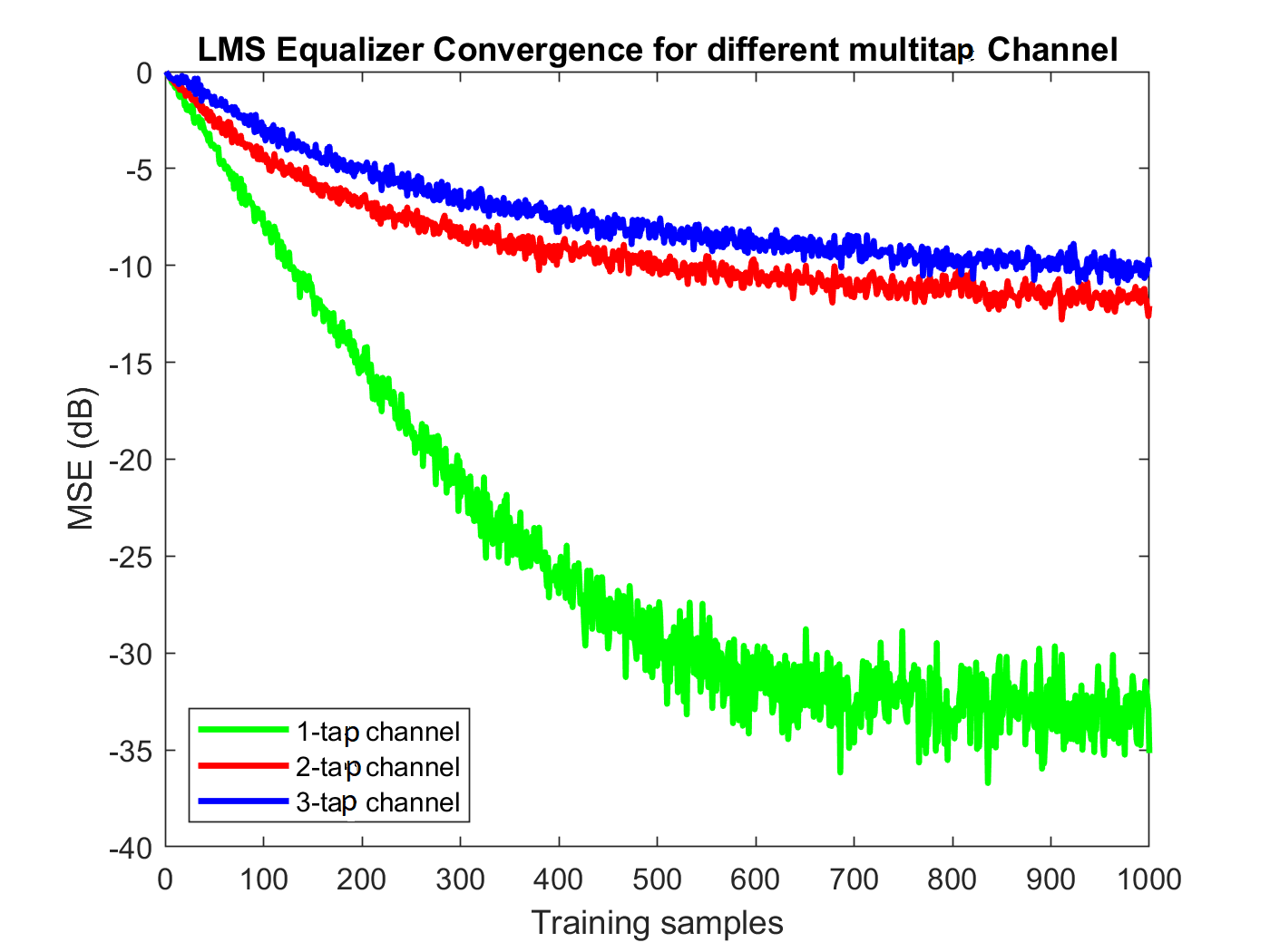}
\caption{LMS equalizer Convergence for flat vs. ISI channel}
\label{fig5}
\end{figure}

Figure \ref{fig4} shows BER performance of single-tap versus multiple-tap LMS channel equalization with 16 taps. The equalizer performance in a flat $\alpha-\mu$ is significantly better than in ISI channel. This is clearly exhibited by algorithm convergence curves for both flat and ISI channels shown in figure \ref{fig5}. 

While LMS algorithm is anticipated to yield better BER performance for multi-tap channels, the BPSK training symbols employed are not whitened. This indicates that the equalizer is selectively canceling channel frequency response under certain frequencies. Since training length is constrained by channel coherence time, whitened training symbols may be used to optimize performance.

\subsubsection{Impact of Training Length}
\begin{figure}[htb] 
\centering
\includegraphics[width = 9 cm,height = 6.6 cm]{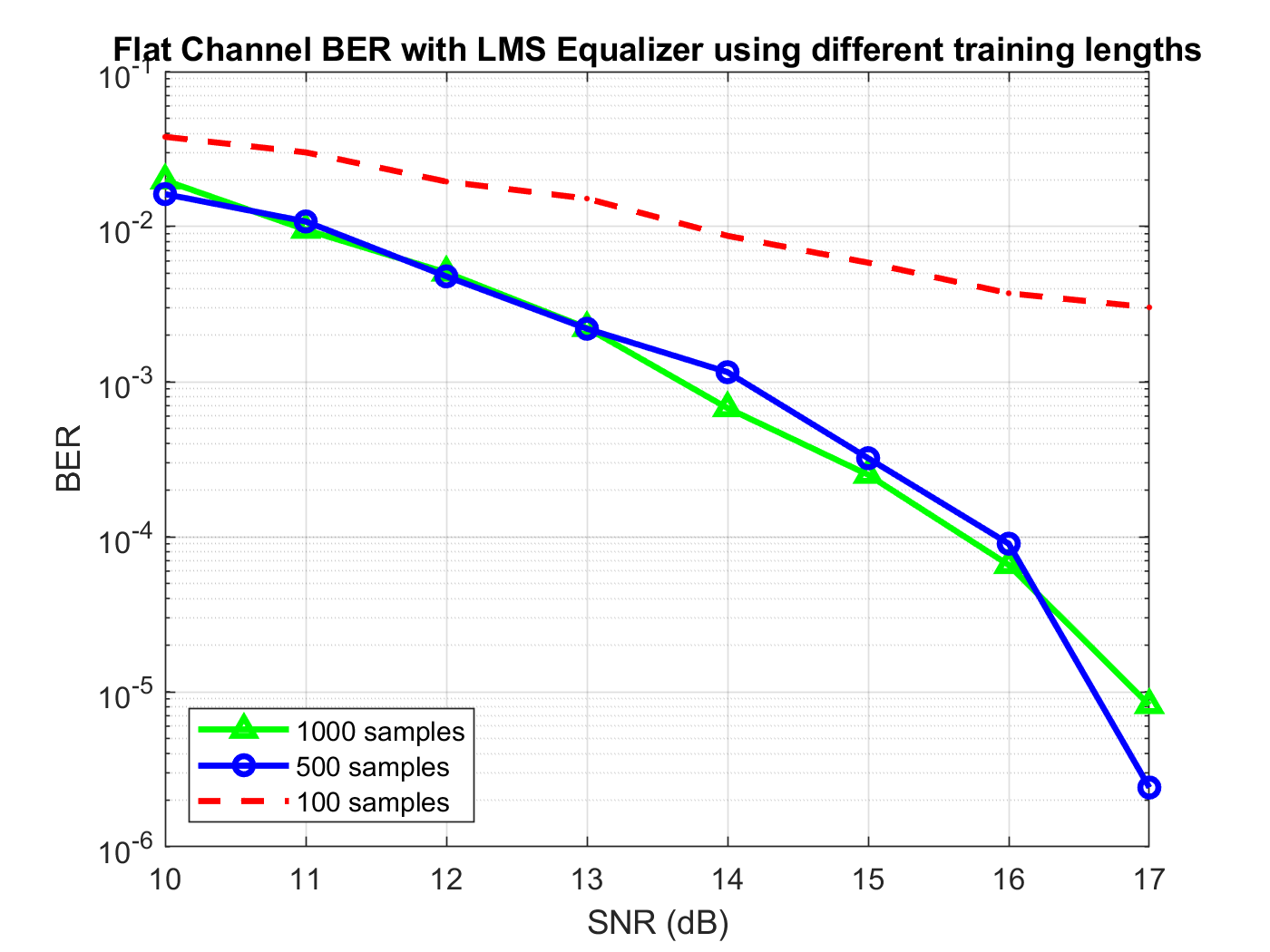}
\caption{LMS equalizer Performance for different training lengths}
\label{fig6}
\end{figure}

Training length determination is dependent on channel coherence time. Since long training length guarantees stable convergence, fast fading channels constraints applicable lengths and thus affect equalizer accuracy. Figure \ref{fig6} illustrates how training length impacts error probabilities. 

It can be shown that employing a training length of about 500-1000 symbols is sufficient for algorithm convergence. A shorter training length of 100 symbols resulted in less effective channel cancellation, proven by high BER values for the specified SNR range. 

\subsubsection{Impact of Equalization Taps}
\begin{figure}[htb] 
\centering
\includegraphics[width = 9 cm,height = 6.6 cm]{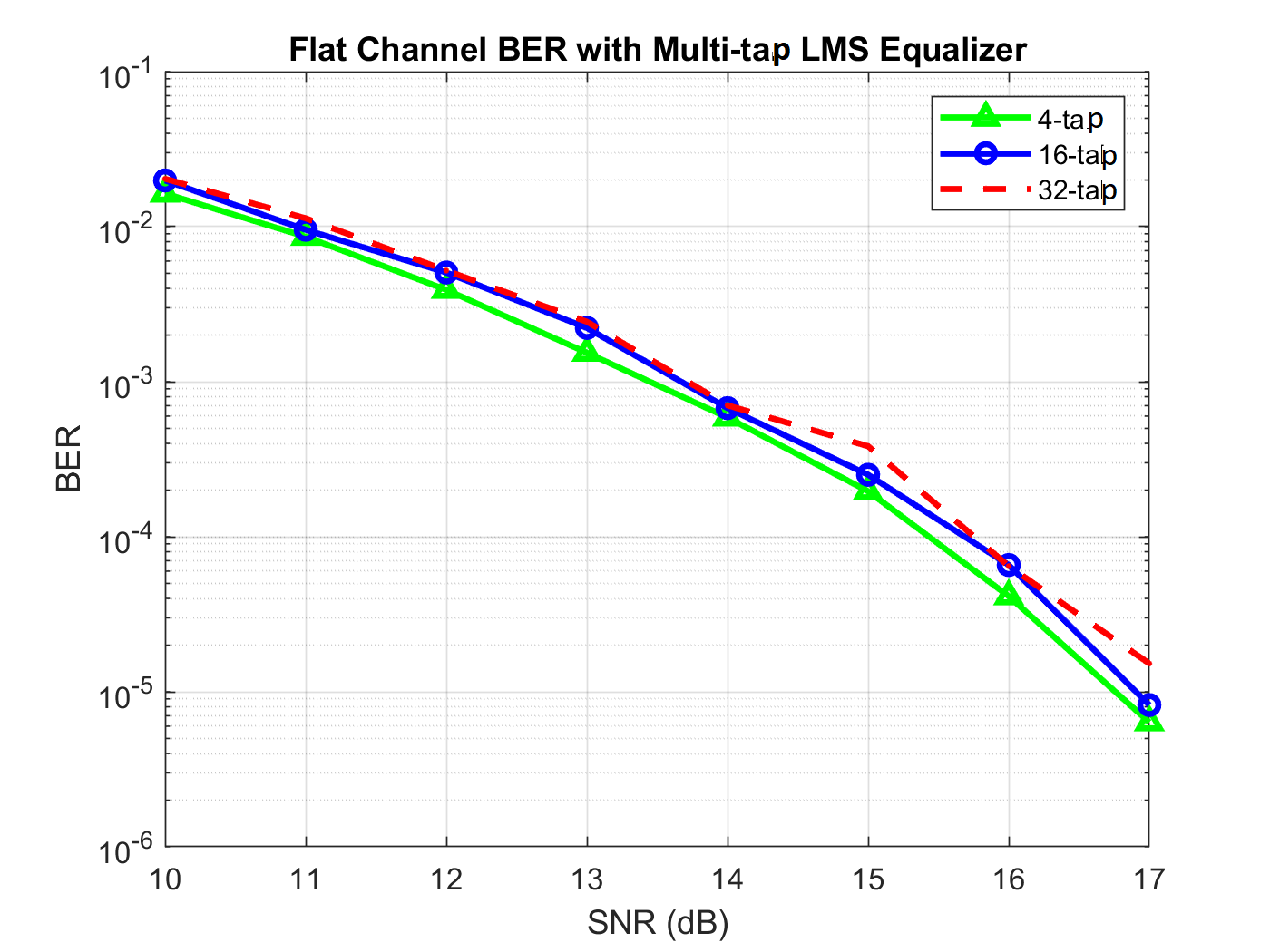}
\caption{LMS equalizer Performance for different numbers of equalizer taps}
\label{fig7}
\end{figure}

The equalizer filter accounts for both ISI channel taps and additive noise effect. While it may seem feasible to adapt equalizer filter with multiple equalization taps more than those of channel, it is yet proven that, for the same convergence rate and training length, the less the number of equalization taps, the better the BER performance. This is exemplified in figure \ref{fig7}. 

\subsubsection{Different parameter setting of the $\alpha-\mu$ channel}
\begin{figure}[htb] 
\centering
\includegraphics[width = 9 cm,height = 6.6 cm]{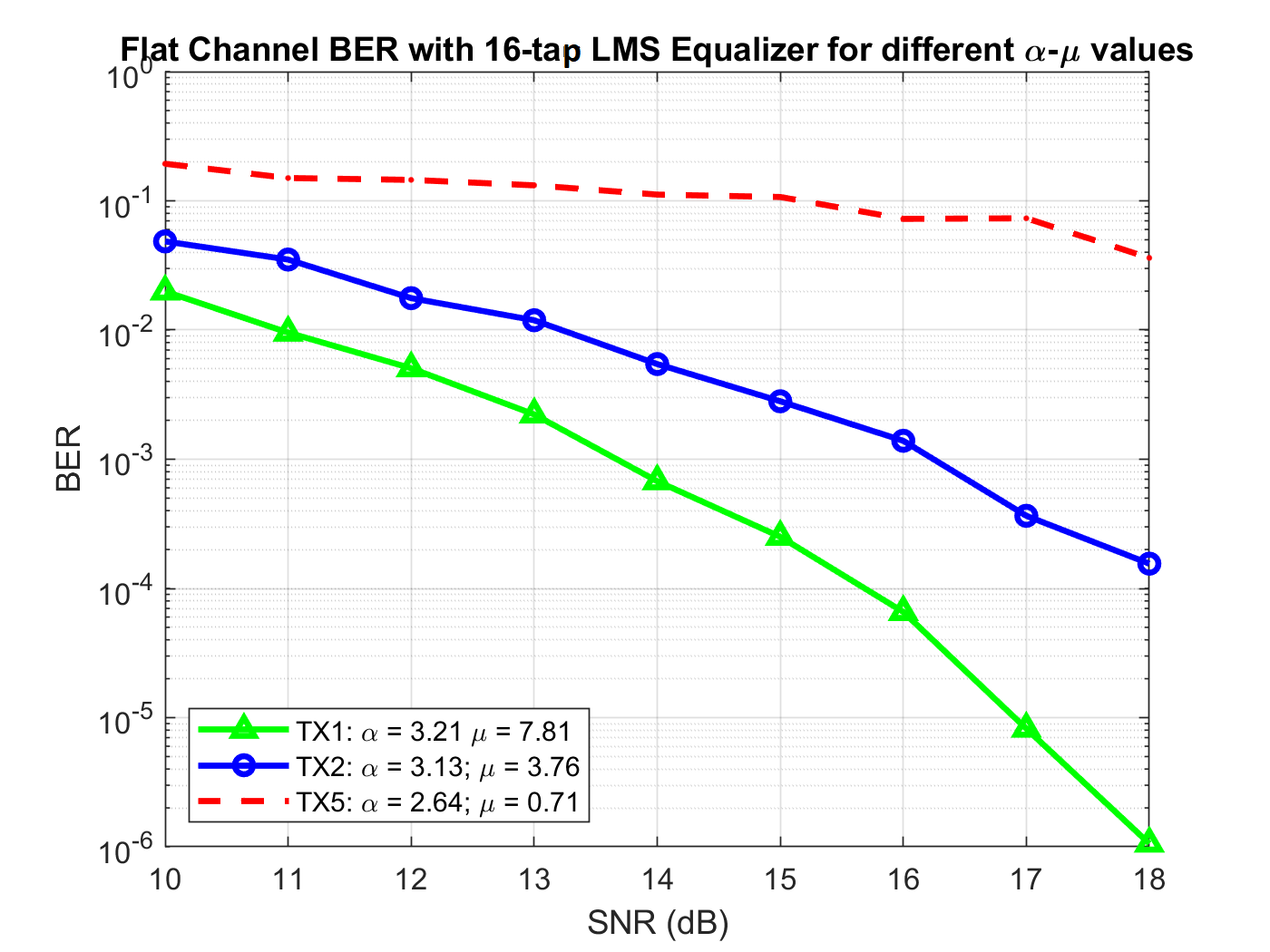}
\caption{LMS equalizer Performance for different $\alpha-\mu$ values}
\label{fig8}
\end{figure}

Assuming a flat fading channel and similar equalization parameters, the state of the channel considerably affects the performance. Different $\alpha-\mu$ values correspond to channel multipath status and fading gain. For a LOS link, lower error probabilities are obtained than non-LOS case. Figure \ref{fig8} addresses this notion through BER curves of the three TX-RX pairs of figure \ref{fig2}. 

\subsection{Recursive-Least-Square Performance}
\begin{figure}[htb] 
\centering
\includegraphics[width = 9 cm,height = 6.6 cm]{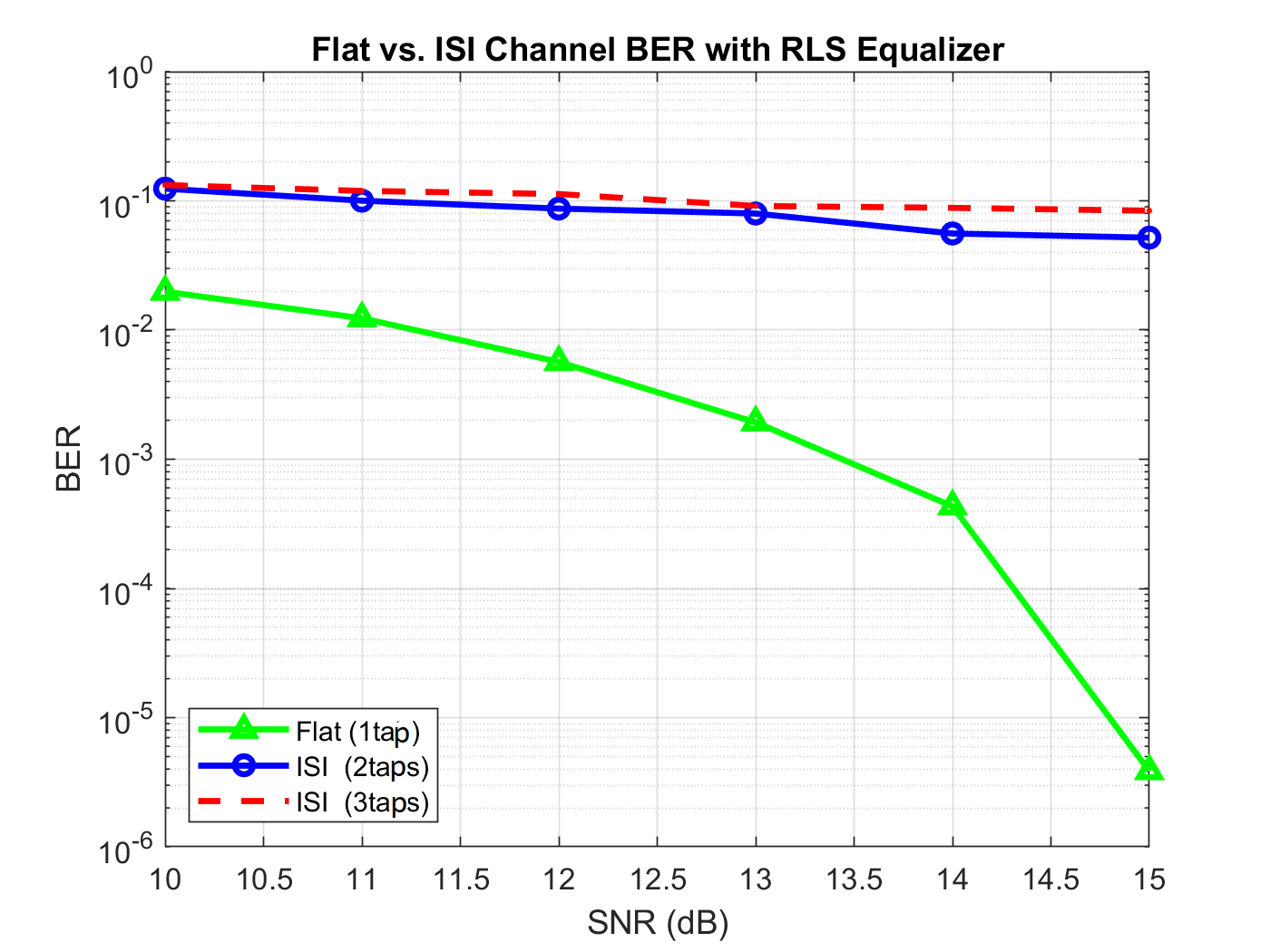}
\caption{RLS equalizer Performance for flat vs. ISI channel}
\label{fig9}
\end{figure}

The RLS algorithm is similar to LMS algorithm in terms of equalizing both channel and noise distortion. Assuming adapted RLS equalizer filter, same remarks are drawn when comparing RLS BER performance for different parameter settings and channel conditions. Figure \ref{fig9} shows RLS BER performance under flat and ISI channels. It is noted that the converged RLS yields similar results as converged LMS.

\begin{figure}[htb] 
\centering
\includegraphics[width = 9 cm,height = 6.6 cm]{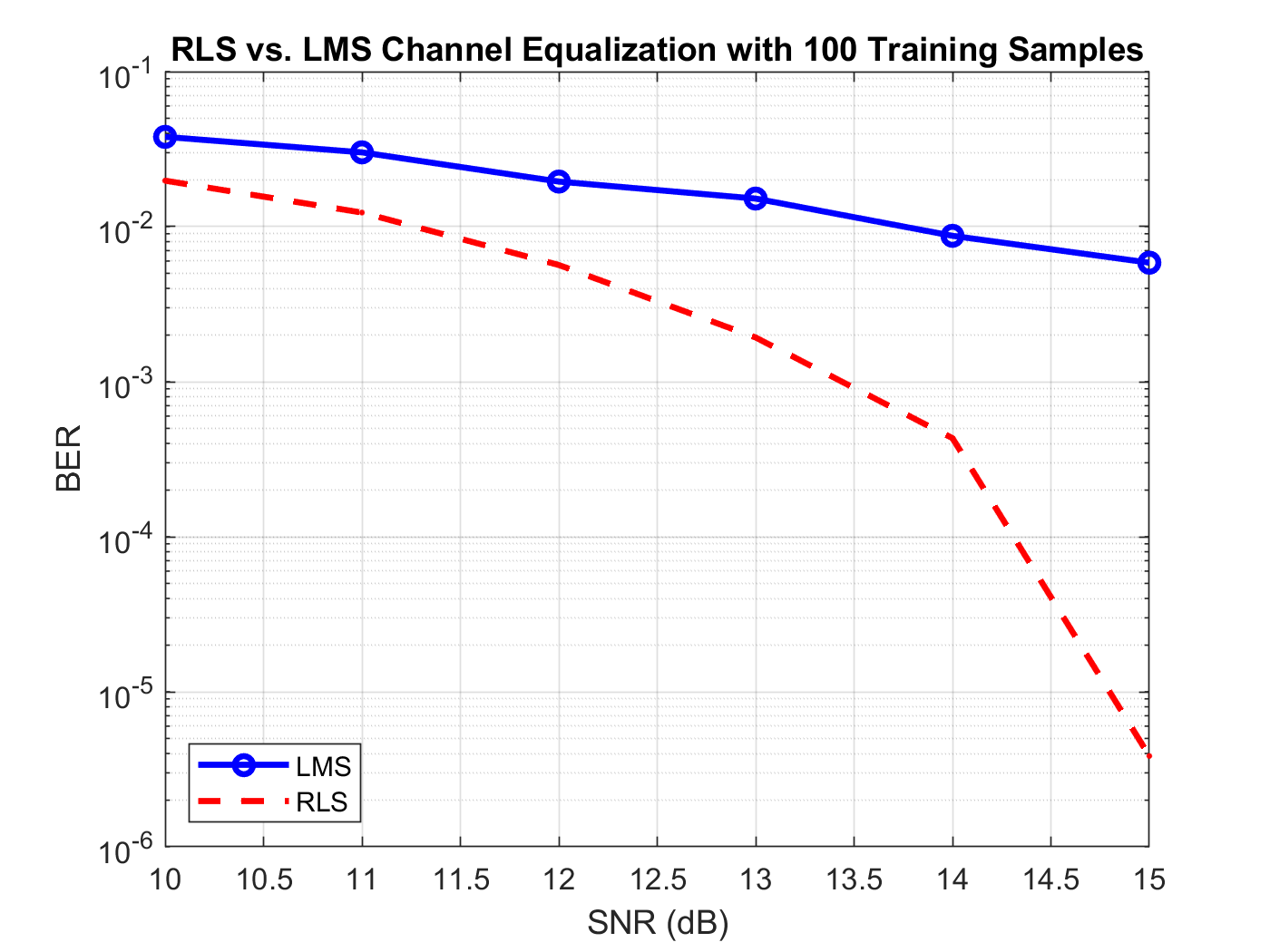}
\caption{RLS vs LMS equalizer Performance for same training length}
\label{fig10}
\end{figure}

\begin{figure}[htb] 
\centering
\includegraphics[width = 9 cm,height = 6.6 cm]{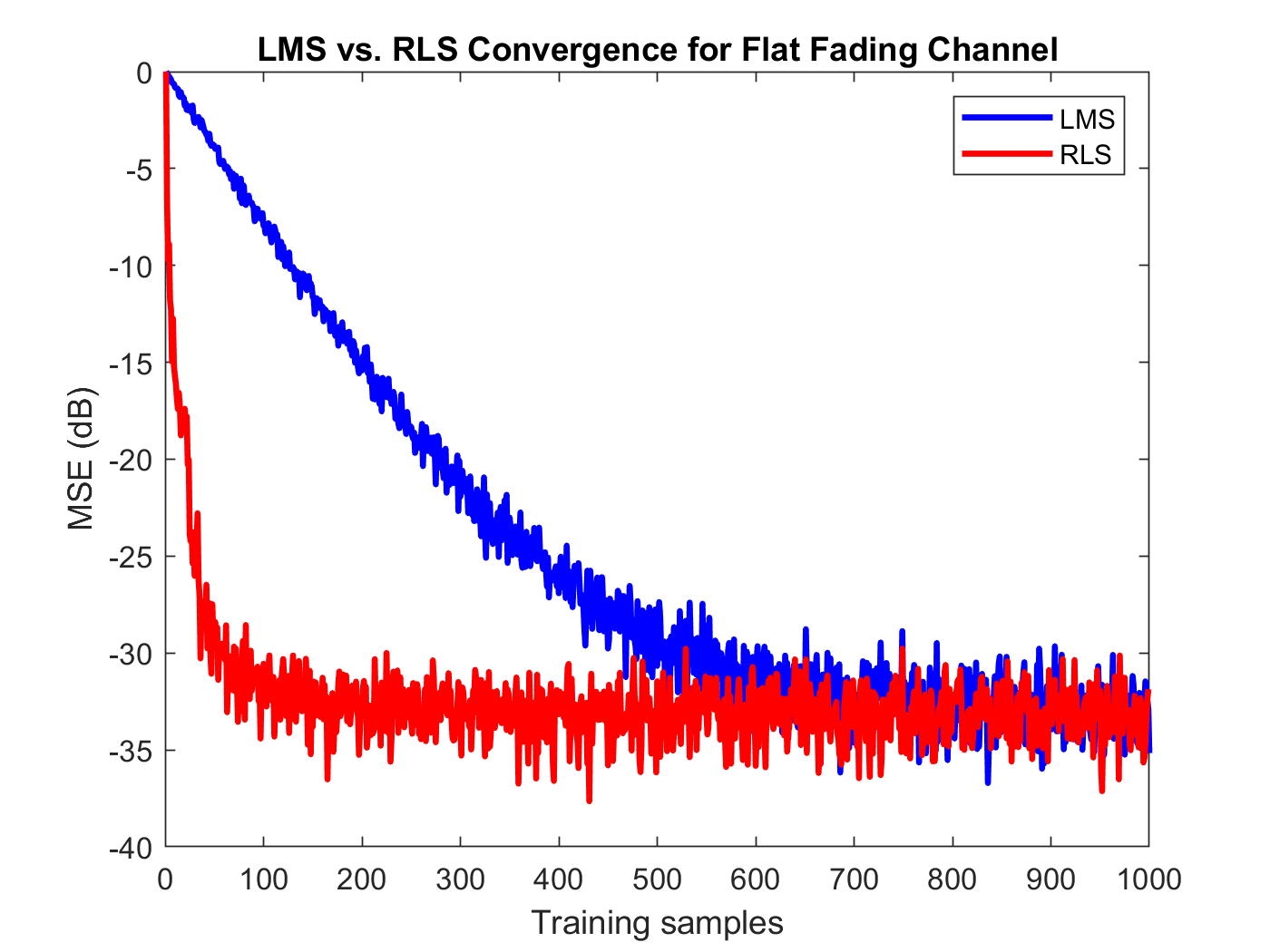}
\caption{RLS vs LMS Convergence for a flat fading channel}
\label{fig11}
\end{figure}

The major advantage of RLS algorithm is significantly fast convergence speed compared to LMS algorithm. This comes at the expense of higher complexity. That is, for short training lengths suited for channel with short coherence time, RLS algorithm results in much reduced error probability. This can be shown in figure \ref{fig10} where a training length of 100 is used for both RLS and LMS. 
Figure  \ref{fig10} result can further be elaborated by addressing RLS versus LMS convergence for the same training length as illustrated in figure  \ref{fig11}. The RLS algorithm required 10 times fewer samples to converge into optimal filter parameters as opposed to LMS convergence. Therefore, under less tight latency constraints, RLS based equalization is much more efficient than that of LMS.
\section{Conclusion}\label{sec:conclusion}

This work considered evaluating channel equalization of multipath wireless channels following $\alpha-\mu$ fading distribution. The ZF, LMS and RLS equalization algorithms were analyzed based on error probability, convergence and implementation complexity. The three algorithms exhibited better performance for flat fading channels compared to ISI channels. Non-adaptive ZF is only applicable when channel effect is dominant compared to noise. Adaptive LMS is studied in terms of suitable training lengths, equalization taps and response to $\alpha-\mu$ parameters variation. It was proven that LMS and RLS algorithms demonstrate similar post-convergence performance. Though LMS algorithm is simpler, it converges much slower than RLS. Fast converging RLS makes it a good fit for rapidly fading channels, with the consideration of high complexity and restricted stability.     

\bibliography{references}
\bibliographystyle{IEEEtran}

\end{document}